\documentstyle[aps,prl,multicol,epsf]{revtex}

\begin{document}

\draft

\title{Nonequilibrium roughening 
transition in a simple model 
of fungal growth 
in $1+1$ dimensions
}

\author{Juan M. L\'opez
and Henrik Jeldtoft Jensen }

\address{Department of Mathematics, Imperial College, 
180 Queen's Gate, London SW7 2BZ, United Kingdom}

\maketitle

\begin{abstract}
We introduce a simple model of yeast-like growth 
of fungi colonies, which exhibits a continuous roughening 
transition far from equilibrium from a smooth ($\alpha = 0$) 
to rough phase ($\alpha = 1/2$) in $1+1$ dimensions. At the 
transition some scaling exponents are calculated by mapping 
the problem onto a directed percolation process. The model 
reproduces the roughening transition observed in some  
experiments of fungal growth. 
\end{abstract}


\begin{multicols}{2}
\narrowtext

Much effort has been devoted to the search for the basic principles
governing the pattern formation
in living organisms. 
Among all the phenomena of formation and development
of complex structures involving living organisms, 
the growth of bacteria and fungi colonies 
has attracted a considerable amount of work in recent years
\cite{bio2,cooper,ben-jacob-rep,ben-jacob1,ben-jacob2,kessler,vicsek,japan1,japan2,japan3,wakita,sams}.

Besides the formation of patterns, which is associated with
the existence of some unstable modes, 
spatio-temporal scale invariance is also commonly observed
depending on the environmental conditions 
\cite{vicsek,japan1,japan2,japan3}.
In the particular case of fungi colonies, the morphology may be
well classified in hyphal and yeast-like growth \cite{carlile}. 
The former corresponds to multicellular growth and 
fractal filamentous patterns
form due to the existence of highly cooperative behavior of
the individual cells. However, yeast-like
growth occurs in solidified media and in this case
the colony is a very compact object. 
The front of the colony usually becomes 
rough ({\it i.e.} the interface width 
diverges with the linear size of the system)
and its dynamics is completely characterized by the
critical exponents, which describe the
scaling properties of the interface fluctuations \cite{alb}.

In recent experiments \cite{sams} 
with the yeast {\em Pichia membranaefaciens}
on solidified agarose film, several 
morphological transitions in $1+1$
dimensions have been reported. In these experiments 
different front morphologies were obtained depending 
on the concentration of polluting 
metabolites. Also transitions from rough to 
flat interfaces were observed
\cite{sams-priv}, although never studied 
systematicly. In the case of {\em Bacillus subtilis}, 
the existence of completely smooth and Eden-like morphologies
(among others) depends on the agar hardness and 
nutrients concentration \cite{wakita}.
Our attention in this Letter is focused 
on this {\em roughening} transition.

In $1+1$ dimensions a surface 
under equilibrium conditions
is rough at any finite temperature  
, because thermal fluctuations 
always make the flat surface entropically unfavorable.  
Only in higher dimensions can equilibrium interfaces
undergo a phase transition from a rough to a smooth interface at some
critical temperature. 
The situation in nonequilibrium is much richer 
because detailed balance is not required and growth processes 
may exhibit a roughening
transition even in $1+1$ dimensions, 
although examples are few \cite{ziff,kertesz,alon}. 
More generally, phase transitions in 
nonequilibrium $1+1$ dimensional systems have usually 
been observed in systems 
with absorbing states. Thus, it is of great interest to find 
models far from equilibrium which do not possess 
absorbing states but still display a phase transition. 

In this Letter we introduce a new class of models 
directly inspired by yeast-like fungal growth 
in $1+1$ dimensions. The model has no absorbing states and
exhibits a nonequilibrium 
roughening transition between a flat and a rough phase. 
Some scaling properties of the model at the transition 
can be related to directed percolation.
We obtain that 
close to the transition a diverging correlation length 
$\xi \sim |\epsilon | ^{-\nu}$ appears, 
where $\epsilon$ is the distance to the critical point, 
and the critical exponent 
is in good agreement with directed percolation in $1+1$ dimensions,
$\nu \approx \nu_{\perp} \approx 1.10$. 
We find that, in the rough phase, 
the model belongs to the Edwards-Wilkinson \cite{ew}
universality class. 

{\em Model.--} We study a class of models motivated
by the growth of colonial organisms like fungi and bacteria.
The model is defined on a 
$1+1$ dimensional lattice in which every site can have two different
states: {\em occupied} or {\em vacant}. Growth of the colony occurs
because of the division of individual cells, so only nearest neighbors
of occupied sites can potentially become occupied. 
The basic idea now is that cell
division is less likely in young cells, so the probability for a vacant site
$i$ of being occupied in the next time step has to increase with 
the {\em total age} $A_i(t)$ 
of the occupied nearest neighboring sites of that vacant site $i$.
At every time step a {\em growth probability} 
\begin{equation}
\label{grow-prob}
P_i(t) = F (\theta \eta_i A_i(t))
\end{equation}
is assigned to all vacant sites that are 
nearest neighbors of occupied sites (note that 
$A_i = 0$ at the remaining  
vacant sites). $F(x)$ is any 
monotonous increasing function 
that satisfies $F(0) = 0$ and $F(\infty) = 1$.
For definiteness, in our simulations we simply used 
$F(x) = {\rm tanh} (x)$.
$\theta$ is an external parameter which 
controls the growth rate. 
The variation in the aging times among the individual cells in the colony
is modelled by the parameter $\eta_i$. 
This random variable is quenched, uncorrelated and uniformly
distributed in $[1-\Delta,1+\Delta]$. 
The simulations are performed on a triangular lattice with 
open boundary conditions, parallelly updating all
the sites according to the growth probability (\ref{grow-prob}). 
The model produces a very compact bulk and
irrelevant overhangs at the scale of the lattice spacing 
may appear on the growth front
depending on the tunning
parameter $\theta$. 

From the point of view of the biological process, this 
model has to be considered as a simplified version of a more realistic
description. The experiments mentioned above 
\cite{sams} have shown that 
yeast-like growth of fungi 
strongly depends on the concentration of nutrients and
inhibitors in the medium. 
In a more complicated model \cite{inh-paper}
the effect of 
inhibitors can be studied by including an inhibitors diffusive field 
$c_i(t)$ which is coupled with the growth probability at every site.
The probability for a vacant site $i$ of becoming occupied is then 
given by $P_i(t) = F(\theta \eta_i A_i(t) 
exp(-c_i(t)/c_0))$, where $c_0$ is
a threshold concentration above which the effect of inhibitors begins to 
be relevant. The model we are presenting here corresponds 
to the limit $c_0 \to \infty$.

{\em Continuous Roughening Transition.--}
Despite its simplicity the model has a non trivial behavior.
For small values of the parameter $\theta$, the fungus
front generated by the model becomes rough and its dynamics 
is characterized as usual by two kinetic 
roughening exponents \cite{alb}. 
In contrast, for large values of $\theta$, the interface 
becomes flat even
when started from a non smooth 
initial condition. This phase transition occurs for any value of 
the disorder intensity $\Delta$. 
We found that these two
phases are separated by a critical threshold 
which depends on the degree of disorder and is at
$\theta_c = 0.183 \pm 0.003$ for 
the case of no disorder $\Delta=0$. 
In the following we restrict ourselves to the
homogeneous case $\Delta=0$ although similar results 
are found for any $0 \le \Delta < 1$.

Let us first study the rough phase, $\theta \ll \theta_c$, in some detail.
Denoting by $h(x,t)$ the front height at time $t$ and substrate
position $x$,
the interface width in a system of size $L$,
$W(L,t) = \langle \overline {[h(x,t) - \overline h(t)]^2} \rangle^{1/2}$
(where overbar and $\langle\cdots\rangle$ 
denote spatial average and over realizations respectively),
measures the magnitude of the interface fluctuations. 
The width is expected
to satisfy the dynamic scaling ansatz \cite{fv,alb}:
\begin{equation}
\label{FV-globalwidth}
W(L,t) \sim 
\left\{ \begin{array}{lcl}
     t^\beta & {\rm if} & t \ll L^z\\
     L^\alpha     & {\rm if} & t \gg L^z,
\end{array}
\right. 
\end{equation}
where $\alpha$ is the roughness exponent and
$z=\alpha/\beta$. 

We have performed simulations 
in systems of different 
sizes to determine the universality class of the model in the rough phase.
Figure 1 shows that
sufficiently far from the transition ($\theta = 0.01$) the 
scaling behavior of the interface 
is given by Eq.(\ref{FV-globalwidth}). The kinetic roughening exponents
$\alpha = 0.46 \pm 0.05$ and $\beta = 0.24 \pm 0.02$ indicate that the
model belongs to the Edwards-Wilkinson universality class \cite{ew}.

Approaching the roughening transition from the rough phase 
a new diverging length
$\xi \sim \epsilon^{-\nu}$, where $\epsilon = \theta_c - \theta$,
appears and enters the scaling of the width. Simulations 
close to the transition ($\epsilon \to 0^{+}$)
were performed in systems of different sizes. In Fig. 2
results for $W(L,t)$ vs. time 
corresponding to $L=900$ and average over 50 realizations
are plotted.
We found that there exist two different
regimes separated by a crossover time $t_c \sim \epsilon^{-\gamma}$,
which diverges as $\epsilon \to 0^+$.  
Figure 2 shows that , near and below the transition, 
$W(L,t)$ scales as $\epsilon^{-\kappa_F}$ for times
$t \ll t_c \ll L^z$ and $W(L,t) \sim t^\beta \epsilon^{\kappa_R}$ for 
$t_c \ll t \ll L^z$. 
$\kappa_{F}$ ($\kappa_{R}$) 
stands for the scaling exponent in the flat (rough) regime 
of the surface width as the
transition is approached.
For times $t \gg L^z$ the interface gets to
saturation as in (\ref{FV-globalwidth}).

The scaling behavior 
in the intermediate time regime before saturation
suggests the following scaling ansatz for times $t \ll L^z$:
\begin{equation}
\label{critical}
W(L,t,\epsilon) = \epsilon^{- \kappa_F} g(t/t_c),
\end{equation}
where the scaling function is $g(u) \sim {\rm const}$ if $u \ll 1$ and
$g(u) \sim u^\beta$ for $u \gg 1$. The scaling relation 
\begin{equation}
\label{scal-rel}
\kappa_R + \kappa_{F} = \beta \gamma   
\end{equation}
between critical exponents must also be fulfilled.
The inset of Fig. 2 shows the data collapse 
for the exponents $\gamma = 1.73 \pm 0.03$
and $\kappa_F = 0.02$ is consistent 
with Eq.(\ref{critical})
over a range of more than four decades.
Due to the very small value of $\kappa_F$ a logaritmic dependence 
$W \sim log(\epsilon)$ cannot be ruled out. 
Using the scaling relation (\ref{scal-rel}) we find 
$\kappa_{R} \simeq 0.41$, which is to be compared with the
the measured value $\kappa_{R} \simeq 0.42 \pm 0.01$.

As seen in some other $1+1$ dimensional 
growth models exhibiting a roughening 
transition \cite{kertesz,alon},
some scaling properties are related 
to directed percolation (DP) \cite{dp}.
Our model turns out to be a model with a maximal velocity.
Small values of $\theta$ produce small growth rates and the front propagates
with a finite velocity. If the parameter $\theta$ is increased, 
the probability for a site of growing ({\it i.e.} becoming occupied)
is also raised as follows from Eq.(\ref{grow-prob}). 
A given site reaches the maximal velocity 
when its probability of growing at the next 
time step is one and then no further increase of the propagation 
velocity at that site can occur \cite{note_vel}.
The transition from rough to flat surface is a DP transition.
The flat phase corresponds to the active 
DP phase whereas the rough phase
corresponds to the non-active DP phase. 
The sites at the highest level
$h = v_{max} t$ are active sites of a DP process.
Below the transition (rough phase) there are no sites growing at 
maximal velocity and the surface mean 
velocity is less than $v_{max}$. 
Above the transition (flat phase),
on the contrary,  
there is a finite density of sites advancing
at $v_{max}$. At $\theta = \theta_c$ the sites
which grow with maximum velocity constitute
a percolation cluster directed along the growth 
direction,
as depicted in Fig. 3.

The mapping to DP allows the derivation of several scaling
exponents. The correlation length $\xi$ 
is identified with the correlation length
of the DP clusters $\xi_{\perp}$, which is transversal to
the growth direction of the interface, 
and $\nu=\nu_{\perp} \approx 1.10$ in $1+1$ dimensions. 
This implies that 
the crossover time $t_c \sim \xi^{z'}$ 
in Eq.(\ref{critical}) 
is given by $t_c \sim \epsilon^{\nu_{\parallel}}$
using the fact that    
$z' = \nu_{\parallel}/\nu_{\perp}$ for DP.
Our estimation of 
the crossover time exponent $\gamma = 1.73 \pm 0.03$ (see Fig. 2)
is in excellent agreement with 
the DP exponent $\nu_{\parallel} \approx 1.73$. 

{\em Order Parameters.--}
We now define a convenient order parameter for the 
rough phase ($\theta < \theta_c$) 
as $V = v_{max} - v$ with $v=\partial_t \overline h(t)$ 
the mean front velocity.
The variable $V$ measures the fraction of the surface points propagating
at velocity lower than the maximal one. 
$V$ is different from zero only in the rough phase and
its scaling behavior can also be related to properties of the associated
DP process.  
In the rough phase $V$ is given by the inverse of the typical
time that the DP correlations survive $\tau \sim \xi_{\parallel}$, which is
the only characteristic time scale. 
So, $V \sim 1/\tau \sim \epsilon^{\nu_{\parallel}}$ close to the transition.
In our simulations we measured $V \sim \epsilon^{1.746 \pm 0.018}$
, as shown in Fig. 4, in excellent agreement with the simple DP scaling
picture.

As for the flat phase ($\theta > \theta_c$)
the density of sites at the maximal height is an appropriate
order parameter \cite{kertesz}, which corresponds to the density 
of active sites in DP. 
However, more interesting here may be the use of appropriate
order parameters which take into account the symmetries of
the model in order to study the possibility of 
spontaneous symmetry breaking (SSB) in the flat phase. It has been
shown that SSB may take place in nonequilibrium situations under
certain conditions even in $1+1$ dimensions \cite{alon,evans}.

The dynamical rules of our model are invariant under  
translation of an integer number of levels in 
the growth direction. The dynamics
do not favor odd nor even heights. Thus it is
adequate to define a magnetisation-like order parameter
\begin{equation}
M(t) = {1 \over L} \sum_{i=1}^{L}(-1)^{h(i,t)},
\end{equation} 
which is no conserved by the dynamics of the model.
This order parameter can be used to quantify the symmetry 
breaking which may take place in the flat phase \cite{alon}. 

In our simulations we found that in the rough phase 
$\langle M \rangle = 0$ and 
also $\langle | M | \rangle = 0$, 
where $\langle \cdots \rangle$ denotes an average over 
realizations.
In the rough phase the interface explores many height levels
and the contribution to the magnetisation at different sites
average out. On the contrary in the flat phase 
$\langle M \rangle = 0$
but $\langle | M | \rangle \neq 0$ due to the fact that most 
heights are at the highest level. However, SSB was not observed
in our model. For every realization $M(t)$ gets to 
a stationary state consisting of flips between $+| M_s |$ and 
$-| M_s |$ at almost every time step. In Ref.\cite{alon} a growth
model with sequential updates has recently been studied in 
which SSB was found. In that model, SSB in the flat phase 
is related to the fact that 
the interface eventually becomes pinned at a certain
height and the
velocity is actually zero in the thermodynamic limit. 
Only in a finite system will the interface have a finite velocity
which vanishes exponentially with the system size. This ensures
the breaking of the symmetry between odd and even heights in the 
thermodynamic limit. The final interface will be pinned at an
odd or even height, depending on the initial conditions, 
breaking the symmetry of the dynamics. 
On the contrary in our model 
the velocity $v_{max}$ in the flat phase is always finite (even
in the thermodynamic limit $L \to \infty$) not allowing SSB.
As occurs in the model reported in Ref.\cite{kertesz} 
the smooth phase does not survive in our model under sequential
dynamics, due to the random walk nature of sequential updates. 

The behavior of the order parameter of the 
flat phase $\langle | M | \rangle$ 
near the transition gives a new scaling exponent
$\eta$, which describes its
decay as the transition is approached from above 
$\langle | M | \rangle \sim (-\epsilon)^{\eta}$.
Our simulations indicate that $\eta \approx 0.50$ which can be compared
with $\eta = 0.55 \pm 0.05$ previously measured for a different 
model \cite{alon}. 

{\em Conclusions.--}
We have studied a model for fungal growth in which the
likelihood of growth at a given site depends on the 
local environment. Vacant sites surrounded by older cells are 
more likely to become occupied than sites neighboring younger cells.
The model exhibits a roughening
transition in $1+1$ dimensions as the growth rate is increased. 
We have found that
the model belongs to the Edwards-Wilkinson 
universality in the rough phase. 
At the transition some critical exponents 
have been calculated by mapping the problem onto a DP process.
Transition from a completely flat to rough interface 
has been observed by T. Sams {\it et. al} 
in experiments \cite{sams-priv} with colonies of the yeast 
{\em Pichia membranaefaciens} on agarose film 
and by Wakita {\it et. al.} \cite{wakita} in the 
bacterium {\em Bacillus subtilis}.

The authors thank A. Saeed for collaboration at 
the early stages of this work and 
M.J. Carlile, K. Christensen, R. Cuerno and T. Sams
for useful comments. This work has been supported 
by the MEC of the Spanish Government and the European Commission.



\begin{figure}
\centerline{
\epsfxsize=5cm
\epsfbox{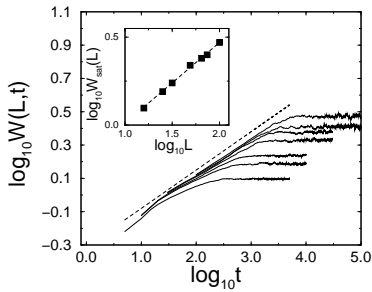}}
\caption{Interface width of the front {\it vs.} time in the rough 
phase ($\theta = 0.01$) for different system sizes 
$L= 16, 25, 32, 50, 64, 75, 100$ . The slope of the dotted line
corresponds to the time exponent $\beta =0.24 \pm 0.02$.
In the inset the values of the width at saturation
are plotted {\it vs.} system size. The dotted line fits
the data and gives the roughness 
exponent $\alpha = 0.46 \pm 0.05$.} 
\end{figure}

\begin{figure}
\centerline{
\epsfxsize=5cm
\epsfbox{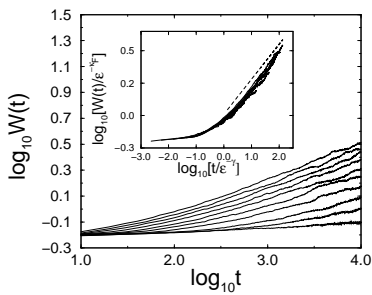}}
\caption{Time behavior of the width 
in the rough phase near the transition. 
Curves correspond to 
different distances to the critical point,
form $\epsilon = 0.083$ (top) to 
$\epsilon = 0.003$ (bottom). 
Data are collapsed in the inset following Eq.(3)
for $\gamma = 1.73$ and $\kappa_F =0.02$, where
the dotted line has slope $0.25$.}
\end{figure}

\begin{figure}
\centerline{
\epsfxsize=5cm
\epsfbox{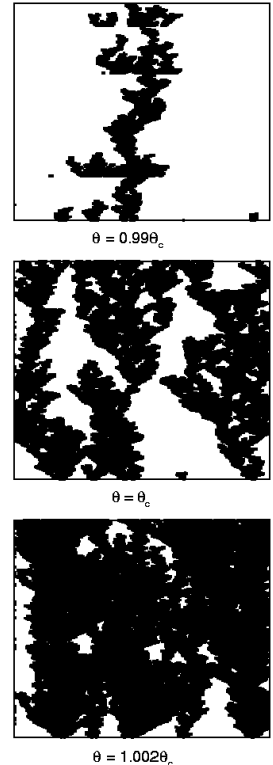}}
\caption{Clusters of sites growing at maximal velocity.
Every pixel represents the position $(x,h)$ of a site
which moved at $v_{max}$ and configurations at intervals
of 5 time steps are shown.
The system size is $L=500$ and time advances upwards.
}
\end{figure}

\begin{figure}
\centerline{
\epsfxsize=5cm
\epsfbox{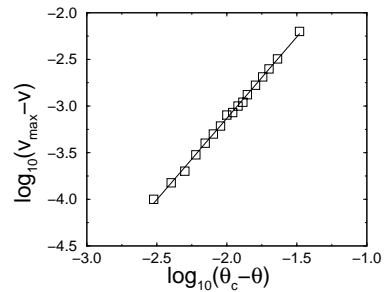}}
\caption{Scaling of the order parameter $V$ in the rough phase near
the transition. The line is a least-square fit of the data and has
a slope $1.746 \pm 0.018$ in good agreement with $\nu_\parallel$.} 
\end{figure}

\end{multicols}


\begin{thebibliography}{99}
\bibitem{note_vel}
The maximal velocity in our model 
is given by one vertical lattice spacing per unit time.
\end{thebibliography}
\end{document}